\documentstyle[11pt,adassconf]{article}  

\begin{document}   

%
%
%

\paperID{P1-26 }

%
%

\title{The VIMOS Mask Preparation Software}

%
%
%

\author{Dario Bottini\altaffilmark{1}, Bianca Garilli}
\affil{Istituto di Fisica Cosmica G. Occhialini, CNR, Milano}

\author{Laurence Tresse}
\affil{Laboratoire d'Astrophysique de Marseille, France}
\altaffiltext{1}{On behalf of the VIRMOS Consortium}

%
%

\contact{Dario Bottini}
\email{bottini@ifctr.mi.cnr.it}

%
%
%

\paindex{Bottini, D.}
\aindex{Garilli, B.}     
\aindex{Tresse, L.}     
%
%



\begin{abstract}          

The main scientific characteristic of VIMOS (VIsible Multi Object
Spectrograph, to be placed on ESO Very Large Telescope) is its high
multiplexing capability, allowing astronomers to obtain up to 800
spectra per exposure. To fully exploit such a potential a dedicated
tool, the VIMOS Mask Preparation Software (MPS), has been designed.
The MPS provides the astronomer with tools for the selection of the
objects to be spectroscopically observed, including interactive object
selection, handling of curved slits, and an algorithm for automatic
slit positioning that derives the most effective solution in terms of
number of object selected per field.  The slit positioning algorithm
must take into account both the initial list of user's preferred
objects, and the constraints imposed either by the instrument
characteristics or by the requirement of having easily reduceable
data.  The number of possible slit combinations in a field is in any
case very high ($10^{73}$), and the task of slits maximization cannot
be solved through a purely combinational approach. We have introduced
an innovative approach, based on the analysis of the function (Number
of slits)/(slit length) vs. (slit length). The algorithm has been
fully tested with good results, and it will be distributed to the
astronomical community for the observation preparation phase.

\end{abstract}

%
%

\section{Introduction}

The VIRMOS (Visible and Infrared Multi-Object Spectrograph) project
consists of two spectrographs with enhanced survey capabilities to be
installed on two unit telescopes of ESO Very Large Telescope (Chile):
VIMOS (0.37-1 $\mu m$) and NIRMOS (0.9-1.85 $\mu m$), each one having
a large field of view (~14'x16') split into 4 quadrants and a high
multiplexing factor (up to approximately 800 spectra per exposure).

To easily exploit such a potential a dedicated tool, the VIRMOS Mask
Preparation Software (MPS), has been implemented.  It provides the
astronomer with tools for the selection of the objects to be
spectroscopically observed, and for automatic slit positioning.  The
output of MPS is used to build the slit masks to be mounted in the
instrument for the spectroscopic observations.

\section{Requirements}

At a limiting magnitude $_I < 24$, the density of objects in the
sky is such that more than 1000 galaxies are visible in a VIMOS
quadrant. Of course, not all these objects can be spectroscopically
observed, as some requirements imposed by data quality have to be
taken into account when placing slits: the minimum slit length will
depend on the object size, since the slit must contain some area of
``pure sky'' to allow for a reliable sky subtraction; spectra must not
overlap either along the dispersion or the spatial direction; as each
first order spectrum is coupled with a second order spectrum which
will contaminate the first order spectrum of the slit above, a good
sky subtraction can be performed only if slits are aligned in columns
(same spatial coordinate) and, within the same column, have the same
length.  All these factor lead to a theoretical maximum number of
spectra per quadrant of approximately 200.

Another requirement for MPS is set by the very good VLT seeing which
allows the use of slits widths of 0.3-0.4 arcsec. Such narrow slits
imply an extremely precise slit positioning, with maximum
uncertainties of the order of 0.1 arcsec. Thus the need for some (1-2
per quadrant) manually selected reference objects (possibly bright and
point-like) to be used for mask alignment.  Moreover, the user must
have the possibility to manually choose some particularly interesting
sources to be included (Compulsory objects) and some others to be
excluded (Forbidden objects) from the spectroscopic sample.  Also a
tool for manual definition of curved or tilted slits, to better follow
the shape of particularly interesting objects, has to be provided.

The MPS works starting from a VIMOS image, to which a catalogue of
objects is associated.  The catalogue can be derived from the image
itself or from some other astronomical data-set.  In this second case,
a way to correlate the celestial coordinates of the objects in the
catalogue with image coordinates is to be provided.  Some catalog
handling capabilities, to allow for the selection of classes of sources
among which to operate the choice of spectroscopic targets, some image
display and catalogue overlay capabilities have to be provided by the
package.

\section{MPS Graphical User's Interface}

As MPS will be distributed to the astronomical community, it should be
based on some already known package (Not Yet Another System).  It was
therefore decided to base the MPS GUI on the SKYCAT tool distributed
by ESO (see \verb1http://archive.eso.org:8080/skycat/1).  This tool 
allows astronomers to couple VIMOS images and catalogues on which to 
operate selections of objects over which to place slits.

A new panel for catalogue display, for object selection (Reference,
Compulsory, Forbidden object) has been implemented.  For each type of
catalogue objects a different overlay symbol has been defined.

\subsection{Curved/tilted slits definition}

A dedicated zoom panel allows the definition of curved/tilted slits.
Curved slits are defined by fitting a Bezier curve to a set of points
chosen by the user by clicking on the zoom display. The fitted curve
is then automatically plotted.  The slit width is chosen through a scale
widget.

Tilted slits can be defined as curved slits and then straightened.  If
the astronomer wants to have slits of a width different from the one
chosen for the automatic slit placements, he can define them as tilted
slits and then align them to the other automatically placed slits.

\section{Slit Positioning Optimization Code}

The core of Mask Preparation Software is the Slit Positioning
Optimization Code (SPOC). Given a catalog of objects, SPOC maximizes
the number of observable objects in a single exposure and computes the
corresponding slit positions.

SPOC places slits on the field of view taking into account: special
objects (reference, compulsory, forbidden), special slits (curved,
tilted or user's dimension defined), spectral first order
superposition, spectral higher order superposition and sky region
parameter (the minimum amount of sky to be added to an object size
when defining a slit).

\subsection{The Optimization}

The issue to be solved is a combinatory computational problem.
Because of the constraint of slits aligned in the dispersion
direction, the problem can be slightly simplified: the quadrant area
can be considered as a sum of strips which are not necessarily of the
same width in the spatial direction. Slits within the same strip have
the same length and the alignment of orders is fully ensured.  The
problem is thus reduced to be mono-dimensional.  It is easy to show
that the number of combination is roughly given by: $N_{combination} =
(N_{possible~ strip~ widths})^{(average~ number~ of~ strips)}$ The
slit length (or strip width) can vary from a minimum of 4 arcsec (20
pixels, i.e. twice the minimum sky region required for the sky
subtraction) to a maximum of 30 arcsec (150 pixels, limit imposed by
the slit laser cutting machine).  The average number of strips can be
estimated as the spatial direction size of the FOV divided by the most
probable slit length: assuming the latter to be 50 pixels (10 arcsec),
we would have $2048/50 = 41$ strips. The number of combinations would
then be: $N_{combinations} = 130^{41} \simeq 4.7\times10^{86}$.
Computing these many combinations would correspond to $10^{60}$ years
of CPU work!  The problem is similar to the well known traveling
salesman problem: in the standard approach, this is solved by randomly
extracting a "reasonable" number of combinations and maximizing over
this subsample. In our case, due to computational time, the
"reasonable" number of combinations cannot be higher than $10^8-10^9$,
so small with respect to the total number of combination that the
result is not guaranteed to be near the real maximum. Our approach has
been to consider only the most "probable" combinations, i.e. the ones
that have the highest probability to maximize the solution.

Step 1: For each spatial coordinate, we can vary the strip width from
the given minimum to the given maximum, count how many objects we can
place in the strip, and build the diagram of the number of slits in a
strip divided by the strip width as a function of the strip width. For
each spatial coordinate, only the strip widths corresponding to peaks
in this histogram are worth considering, as they correspond to local
maxima of the number of slits per strip.  The exact positioning of the
peaks varies for each spatial coordinate, but the shape of the
function remains the same. The position of the peaks can be easily
found in no more than 6-7 trials (using a partition exchange method).

Step 2 For each spatial coordinate we have K (where K is the number of
peaks) possible strips, each with its own length and number of slits.
Although the number of combinations to be tested is decreased it is
still too big in terms of computational time.  

Step 3 A further reduction can be obtained if, instead of considering
all the strips simultaneously, we consider sequentially M subsets of N
consecutive strips, which together cover the whole FOV. At this point,
we should vary N (and consequently M) to find the best solution. In
practice, when N is higher than 8-10, nothing changes in terms of
number of observable objects. For N=10, thus M=4
(i.e. $2048/(4\times50)$), the number of combination is reduced to
only $4\times4^{10}=4\times10^6$, which means a few seconds of CPU
work.  Unfortunately, as a consequence of the optimization process,
small size objects are favored against the big ones.

\subsection{An alternative optimization}

A second, less optimized algorithm, has been implemented within SPOC.
This alternative algorithm does not optimize all strips simultaneously
but builds the $N_{Slit}/Strip$ function strip by strip without
considering object sizes, and takes only the maximum of the
distribution. Then it enlarges each strip width by taking into account
object sizes.  In this way the number of placed slits decreases by a
few percent but the object dimension bias disappears.

\section{SPOC Graphical User Interface}

A dedicated panel for SPOC set up has been implemented within SKYCAT.
Trough this panel, users can select the grism, the slit width, and the
sky region parameter, the number of masks to be obtained for the given
field, and the type of SPOC maximization.
 
The number of input slit and placed slit for all kinds of objects
(Reference, Compulsory, etc...)  is printed in a text box.

The slit catalogue produced by SPOC can be loaded as a normal SKYCAT
catalog with overlay symbols defined for all kinds of objects, and it
is also possible to plot the slit and spectrum overlay for all SPOC
catalog objects.

\end{document}